\documentclass[%
 reprint,
 amsmath,amssymb,
 aps,
]{revtex4-2}

\usepackage{graphicx}% Include figure files
\usepackage{dcolumn}% Align table columns on decimal point
\usepackage{bm}

\usepackage{epstopdf}
\AppendGraphicsExtensions{.tif}

\begin{document}

\title{Mechanical properties of single and polycrystalline solids from machine learning}

\author{Faridun N. Jalolov}
\affiliation{Skolkovo Institute of Science and Technology, Skolkovo Innovation Center, Bolshoy Boulevard 30, bld. 1, Moscow 121205, Russia}

\author{Evgeny V. Podryabinkin}
\affiliation{Skolkovo Institute of Science and Technology, Skolkovo Innovation Center, Bolshoy Boulevard 30, bld. 1, Moscow 121205, Russia}

\author{Artem R. Oganov}
\affiliation{Skolkovo Institute of Science and Technology, Skolkovo Innovation Center, Bolshoy Boulevard 30, bld. 1, Moscow 121205, Russia}

\author{Alexander V. Shapeev}
\affiliation{Skolkovo Institute of Science and Technology, Skolkovo Innovation Center, Bolshoy Boulevard 30, bld. 1, Moscow 121205, Russia}

\author{Alexander G. Kvashnin*}
\affiliation{Skolkovo Institute of Science and Technology, Skolkovo Innovation Center, Bolshoy Boulevard 30, bld. 1, Moscow 121205, Russia}

\email{A.Kvashnin@skoltech.ru}

\date{\today}

%%%%%%%%%%%%%%%%%%
\begin{abstract}
Calculations of elastic and mechanical characteristics of non-crystalline solids are challenging due to high computation cost of $ab$ $initio$ methods and low accuracy of empirical potentials.
We propose a computational technique towards efficient calculations of mechanical properties of polycrystals, composites, and multi-phase systems from atomistic simulation with high accuracy and reasonable computational cost.
It is based on using actively learned machine learning interatomic potentials (MLIPs) trained on a local fragments of the polycrystalline system for which forces, stresses and energies are computed by using $ab$ $initio$ calculations.
Developed approach is used for calculation the dependence of elastic moduli of polycrystalline diamond on the grain size.
This technique allows one to perform large-scale calculations of mechanical properties of complex solids of various compositions and structures with high accuracy making the transition from ideal (single crystal) systems to more realistic ones.
\end{abstract}

\maketitle

%%%%%%%%%%%%%%%%%%
\section{Introduction}
\label{sec:intro}
%%%%%%%%%%%%%%%%%%

Diamond is widely used material due to its unique properties and, first of all, its unsurpassed hardness (varying from 60 to 120 GPa \cite{kvashnin2019computational, andrievski2001superhard, field2012mechanical, blank1998ultrahard} depending on conditions) attracting a constant demand in the manufacturing industry.
Synthetic diamonds, which are mainly used in industry, usually synthesized in a polycrystalline structure.
Depending on the method of production and parameters of the technological process the size of crystallites (grains) of such diamonds may vary from a few nanometers to tens of microns \cite{scott2018influence}.
The mechanical properties of polycrystalline diamonds depend on the size of the grains \cite{lammer1988mechanical}.
In the case of large grains (about a micron) the specific volume of intergrain boundaries is not large, and the basic mechanical properties of such diamonds are close to those of single crystal.
However, the specific volume of inter-granular boundaries increases with decreasing grain size, which significantly affects the mechanical properties of diamonds.
According to Refs. \cite{huang2014nanotwinned, irifune2003ultrahard} the elastic properties of polycrystalline diamond may even exceed the mechanical properties of single crystal diamond.
Understanding of how the properties of polycrystalline diamond depend on the grain size is important from a practical point of view and taking into account the technologies for synthesis of polycrystalline diamonds from ultrafine diamond dust.
The practical need for comprehensive and accurate theoretical study of the effect of grain size in polycrystalline diamonds on their mechanical properties motivated this work. 

Perhaps the most adequate approach to study this problem is to simulate the system at the atomistic level.
However, a critical aspect of atomistic simulation is the choice of a model of interatomic interaction.
Traditionally, there are two approaches for such models, namely empirical potentials and \textit{ab initio} calculations.
Empirical potentials are used to perform simulations of large atomistic systems because of their computational efficiency.
Such models have a fixed functional form, constructed by insight, and have only several fitting parameters, which are chosen to reproduce the basic properties of single crystals and experimental results in simulation.
The widely used empirical potentials for diamond are Tersoff potential \cite{tersoff1988empirical}, Brenner potential \cite{brenner2002second}, and ReaxFF force field \cite{van2001reaxff}.
However, the accuracy of empirical potentials may not be sufficient to reproduce the complex nature of interactions in the region of inter-granular boundaries, where the structure is different from the regular crystal lattice  where the potentials have not been fitted to.

In the work by Erohin et al. \cite{erohin2015elastic} the nature of ultrahigh hardness of polycrystalline diamonds was theoretically studied by using molecular dynamics simulations with Brenner potential \cite{brenner2002second}.
Authors traced the evolution of the bulk modulus with the grain size and found structures with bulk modulus higher than that of single crystal diamond.
Despite of the fact that description of new atomic configurations in polycrystals by classical empirical potentials is negotiable this study showed an idea that unusually high bulk modulus may caused by anisotropic response of the particular grains to the hydrostatic stress.
This hardening mechanism seems quite plausible in view of agreement with the reference experimental data.

Among the quantum-mechanical methods the most widely used for description of materials properties is the density functional theory (DFT) \cite{sham1966one, hohenberg1964inhomogeneous}.
DFT provides a high-accuracy calculations of the energies and forces, but its practical application is limited to atomistic systems with several hundred atoms, which makes them inapplicable for describing the inter-granular boundaries.

Recently, models of interatomic interaction based on machine learning have been rapidly developing and gaining popularity.
They are designed to combine the computational efficiency of empirical potentials and the accuracy of quantum-mechanical models.
In contrast to empirical potentials, the machine-learning interatomic potentials (MLIPs) have a flexible functional form that allows one to approximate any potential energy surface with a predetermined accuracy (at least theoretically) by increasing the number of parameters.
Nowadays, there are several MLIPs which use different types of representations of crystal structures, such as GAP \cite{bartok2013representing}, MTP \cite{shapeev2016moment}, NNP \cite{behler2007generalized,behler2011neural} etc.
The use of machine learning (ML) techniques in the context of atomistic simulation of materials has gained considerable momentum in the past decade \cite{behler2007generalized, bartok2010gaussian, rupp2012fast, montavon2013machine, noe2020machine, butler2018machine, schutt2018schnet, friederich2021machine, lopanitsyna2021finite,imbalzano2021uncertainty, cheng2019ab, carleo2019machine, dragoni2018achieving, bartok2018machine, isayev2015materials, szlachta2014accuracy, deringer2017machine}.
Generally in the training procedure, the potential parameters are determined from the requirement of minimizing the deviation between the forces and energies predicted and calculated from the first principles on the configurations from the training set.
However, if the atomistic configuration for which the energies and forces are calculated is significantly different from that presented in the training set, extrapolation occurs and the prediction error may be unacceptably high.
To resolve the extrapolation problem, MLIP must recognize the configurations on which the extrapolation will occur.
Efficiently, this procedure can be organized as learning on-the-fly \cite{podryabinkin2019accelerating}.
This scheme \cite{podryabinkin2019accelerating} ensures that there is no extrapolation when calculating the energy of forces for atomistic configurations.

In this work we propose an active learning method for MLIPs with automatic build up of the local configuration fragments on which the potential extrapolates to a periodic configuration with regular periodic joint.
The size of such configurations are small enough, so they are suitable for DFT calculations.
Thus, this work has two aims: (1) to study the dependence of the elastic properties of polycrystalline diamond on the grain size with the accuracy close to DFT, and (2) to test the active learning method on the local environments with their build up to a periodic configurations. 

%%%%%%%%%%%%%%%%%%%%%%%%%%
\section{Methods}
\label{sec:comp}
\subsection{Machine learning interatomic potentials}

The development and dissemination of MLIPs have revolutionized computational materials science.
Application of MLIPs makes it quite easy to solve issues previously considered unsolvable or unreasonable for solving due to enormous resource consumption.
First of all, MLIPs enable solving the problems of simulation of the systems with a large number of atoms, or problems where the calculations of physical properties of a huge number of systems are required to be done in a reasonable time.
In particular, MLIPs enable the calculations of the nanohardness of various materials based on the first principles \cite{podryabinkin2022nanohardness}, high-throughput screening and accelerating crystal structure prediction \cite{gubaev2019accelerating, podryabinkin2019accelerating}, long molecular dynamics simulations \cite{shapeev2020elinvar}. 

In this work we use the Moment Tensor Potentials (MTPs) \cite{shapeev2016moment} as interatomic interaction model.
MTPs belong to the class of local machine-learning potentials, where the total energy of the configuration is composed of contributions $V$ of individual atoms (site energies) as

\begin{equation} \label{eq:E_MTP}
E^{\rm mtp}({\rm cfg}) = \sum_{i=1}^n V(\mathfrak{n}_i).
\end{equation}

The site energy of atom $i$ depends on a local atomic neighborhood $\mathfrak{n}_i = z_i, z_j, \mathbf{r}_{ij}$, which is determined by the type of central atom $z_i$, by the types $z_j$, and relative positions $\mathbf{r}_{ij} = \mathbf{r}_{j}-\mathbf{r}_{i}$ of neighboring atoms within the cutoff radius $\mathbf{r}_{j}-\mathbf{r}_{i} \leq R_{ \rm cut}$.
The site energies $V(\mathfrak{n}_i)$ are calculated as a linear combination of basis functions $B_{\alpha}(\mathfrak{n}_i)$ 

\begin{equation} \label{eq:V_MTP}
V(\mathfrak{n}_i) = \sum \limits_{\alpha} \xi_{\alpha} B_{\alpha}(\mathfrak{n}_i).
\end{equation}

Coefficients $\xi_{\alpha}$ of this linear combination are the subset of parameters of the potential and are found in the training procedure.

Definition of the basis functions is based on the moment tensor descriptors:

\begin{equation} \label{eq:moment}
M_{\mu,\nu}({\mathfrak{n}}_i)=\sum_{j} f_{\mu}(|r_{ij}|,z_i,z_j) \underbrace {\mathbf{r}_{ij}\otimes...\otimes \mathbf{r}_{ij}}_\text{$\nu$ times}.
\end{equation}

Here $\underbrace {\mathbf{r}_{ij}\otimes...\otimes \mathbf{r}_{ij}}_\text{$\nu$ times}$ is a tensor of rank $\nu$, 

\begin{equation}\label{eq:f_mu}
f_{\mu}(|\mathbf{r}_{ij}|,z_i,z_j) = \sum_{\beta=1}^{N_Q} c^{(\beta)}_{\mu, z_i, z_j} Q^{(\beta)} (|r_{ij}|),
\end{equation}

is a scalar radial function, where $\big\{ c^{(\beta)}_{\mu, z_i, z_j} \big\}$ is the set of "radial" parameters, 

\begin{equation} \label{eq:radial-basis}
\displaystyle
Q^{(\beta)}(|r_{ij}|) =\begin{cases}
\varphi^{(\beta)} (|r_{ij}|) (R_{\rm cut} - |r_{ij}|)^2 & |r_{ij}|<R_{\rm cut} \\
0 & |r_{ij}| \geq R_{\rm cut}
\end{cases}
\end{equation}
are the radial basis function $\big\{ c^{(\beta)}_{\mu, z_i, z_j} \big\}$, based on Chebyshev polynomials $\varphi^{(\beta)}$.
$B_{\alpha}(\mathfrak{n}_i)$ are constructed from $M_{\mu,\nu}({\mathfrak{n}}_i)$ as various convolutions of tensors of different ranks yielding a scalar.
In addition to the energy of the configurations, the implementation of the MTP allows the calculation of the forces on atoms and virial stresses of the configuration based on the analytical derivatives of $E$ with respect to the positions of atoms.

The parameters of the radial functions $\big\{ c^{(\beta)}_{\mu, z_i, z_j} \big\}$ together with the linear parameters $\xi_{\alpha}$ form a set of parameters $\mathbf \theta$ of MTP, which are found in the training procedure. 
This procedure minimizes the standard deviation between the energies, forces, and stresses computed by DFT and MTP over a set of configurations (training set):

\begin{widetext}
\label{Fitting}
\begin{eqnarray}{c}
\displaystyle
\sum \limits_{k=1}^K \Bigl[
	w_{\rm e} \left(E^{\rm mtp} ({\rm cfg}_k; {\mathbf {\theta}}) - E^{\rm dft}({\rm cfg}_k) \right)^2
	+
	w_{\rm f} \sum_{i=1}^{N_k} \left| {\mathbf{f}}^{\rm mtp}_i({\rm cfg}_k; {\mathbf {\theta}}) - {\mathbf{ f}}^{\rm dft}_i({\rm cfg}_k) \right|^2 
\\ \displaystyle
	+
	w_{\rm s} \big|\sigma^{\rm mtp}({\rm cfg}_k; \mathbf {\theta}) - \sigma^{{\rm dft}}( {\rm cfg}_k)\big|^2 \Bigr] \to \min\limits_{\mathbf \theta}.
\end{eqnarray}

\end{widetext}

The Newton's method of second order is used as minimization algorithm.

\subsection{Active learning on-the-fly with local atomistic environments}

Probably the main difficulty in using a MLIP is related to their transferability.
Since the calculation of energies and forces by the MLIP can be seen as an interpolation of these quantities over the training set, it is important that the training set covers the domain of the configuration space where the energy and forces are calculated.
Otherwise, extrapolation will occur and such predictions are likely to have very low accuracy.
For example, a MLIP learned only on bulk configurations will extrapolate when calculating energies and forces of a free surface.
Therefore, when using a MLIP, it is important to have a mechanism for recognizing the extrapolations (without performing first-principles calculations), which also referred as active learning methods.
When an extrapolation is recognized, the corresponding configuration can be computed with the DFT and then added to the training set.
The training domain expands and MLIP will not extrapolate further on that configuration.
It should be noted that during MD simulations, the trajectory can go beyond the training set, even if there was no extrapolation at the initial part of the MD trajectory.
Therefore, one of the most efficient method of using MLIP is to make MD simulations with extrapolation control and learning the potential on-the-fly.

\begin{figure*}
  \begin{center}
    \begin{minipage}[h]{1\linewidth}
      \center{\includegraphics[width=1\linewidth]{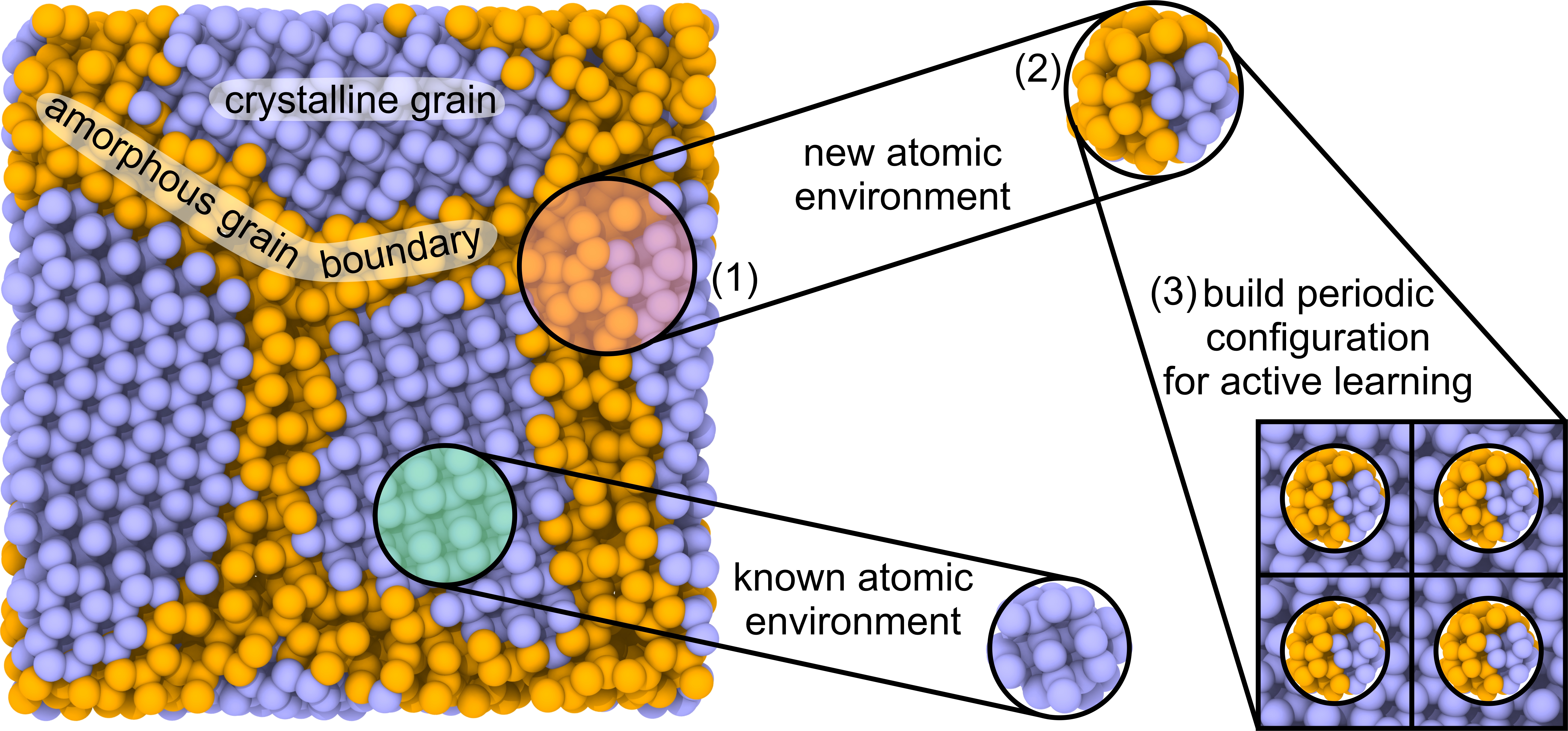}}  \\
    \end{minipage}
  \caption {Schematic illustration of learning on the local atomistic environment. Region highlighted by red (1) contains atoms with highest extrapolative grade, which then cut from the structure (2) and used to build the periodic configuration (3) for further DFT calculations of energy, forces, and stresses.}
  \label{fig:local}
  \end{center}
\end{figure*}

Different MLIPs have their own methods allowing the recognition of extrapolations.
For example, MLIPs based on Gaussian Processes, as such a mechanism, use prediction variation estimation \cite{zuo2020performance}.
Neural network-based MLIPs, detect extrapolation based on monitoring model committee disagreement \cite{schran2020committee}.
For MTPs, the degree of extrapolation is calculated from the principle of maximum volume of the domain in configuration space spanned on the training set and is calculated with MaxVol algorithm \cite{goreinov2010find}.
The degree of extrapolation can be estimated for the whole configuration, as well as for atomistic neighborhoods $\mathfrak{n}$ of individual atoms \cite{podryabinkin2017active}.
The second method allows one to detect local fragments of the configuration with potentially low accuracy of force calculations.
This is especially in demand when working with configurations with a large number of atoms.
However, the problems arise here with the obtaining of \textit{ab initio} data due to the practical impossibility of calculating large configurations with DFT.
This problem can be solved by somehow cutting out the extrapolation fragments from a large configuration, with the  number of atoms suitable for DFT calculations (in practice, usually not more than a couple of hundred of atoms).

In recent papers \cite{podryabinkin2022nanohardness,podryabinkin2023mlip} the extrapolated atomistic environments were simply cut out and further computed as non-periodic atomic clusters.
Such an approach is reasonable when we deal with free surfaces in the simulated system.
However, in the our work only bulk configurations are treated, and training the potential on fragments with a free surfaces will lead to an unreasonable expansion of the training domain to non-relevant areas with subsequent decrease in the accuracy.
Therefore, in this paper we realized another approach based on the construction of periodic configurations from cut fragments.
Namely, this is done as follows.

\begin{enumerate}
\item We identify atomistic environments $\mathfrak{n}$ on which MLIP extrapolates (step (1) in Fig. \ref{fig:local}).

\item From the whole configuration, we cut the atoms inside the cube containing the cutoff sphere with the extrapolative environment $\mathfrak{n}$. 
The size of the cube may be slightly larger then $2R_{\rm cut}$ (step (2) in Fig. \ref{fig:local}).

\item Next we construct a periodic supercell with this cube having cell parameters 0.5 {\AA} larger at each side of the cube than the cut one to avoid appearance of extremely short interatomic distances after applying periodicity (step (3) in Fig. \ref{fig:local}). 

\item In the resulting periodic configuration we relax the lattice vectors and positions of all atoms outside the extrapolation sphere.
The atoms inside the extrapolation sphere remain fixed and do not change their positions, which guarantees that the extrapolative environment does not change during relaxation. 

\end{enumerate}

The relaxation in the last step is carried out in two steps: (1) pairwise repulsive potential is used to fix too short interatomic distances, and (2) DFT for calculations of energies, forces, stresses. 
This essentially constructs the periodic joint similar to a regular inter-granular boundary in the cell, and eliminates the formation of irrelevant atomistic fragments on it. 

%%%%%%%%%%%%%%%%%%%%%%%%%%
 
%%%%%%%%%%%%%%%%%%%%%%%%%%

\section{Computation details}
\subsection{Generation of polycrystalline structure samples}

The very first step in elastic moduli calculation is generation of periodic polycrystalline samples.
For this purpose we use the Voronoi tessellation method \cite{brostow1978construction, finney1979procedure, tanemura1983new} as implemented in Atomsk\cite{hirel2015atomsk}.
The method splits a given periodic domain into specified number of grains with the random shape and orientation.

\begin{figure*}
  \begin{center}
    \begin{minipage}[h!]{1\linewidth}
      \center{\includegraphics[width=1\linewidth]{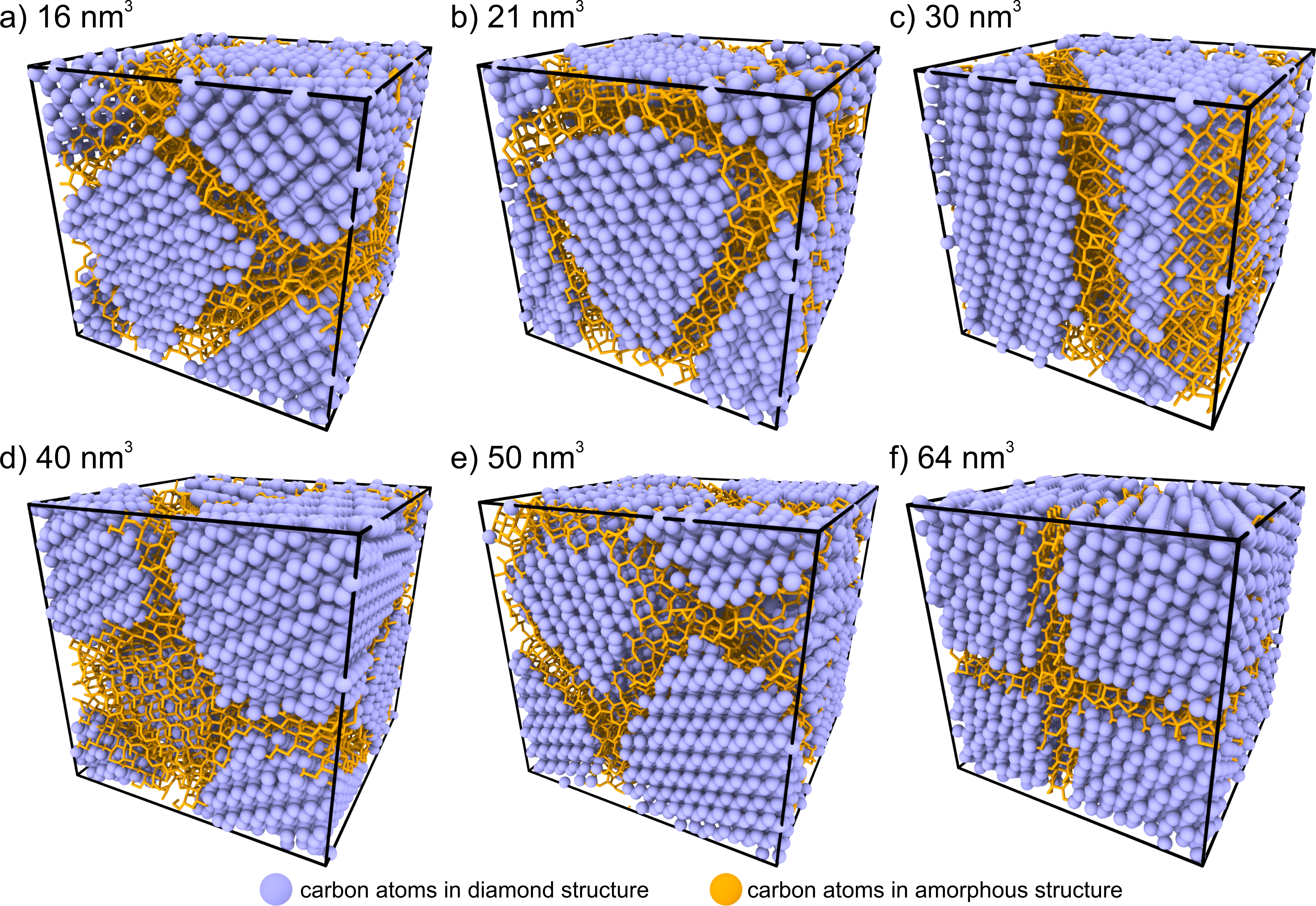}}  \\
    \end{minipage}
  \caption {Crystal structure of polycrystals with different grain volumes of a) 16, b) 21, c) 30, d) 40, e) 50, and f) 64 $nm^3$ generated and considered in our work. By orange and violet colors the carbon atoms in amorphous and diamond structure are shown respectively.}
  \label{fig:polycryst}
  \end{center}
\end{figure*}

The computational domain had a cubic shape with the size $4\times4\times4$ nm. By variation number of grains we generated several diamond polycrystals with different grain sizes. For example, there are 4 grains in polycrystalline with 16 $nm^3$ average grains size and only 1 grain in polycrystalline with 64 $nm^3$ average grains size. To study the dependence of mechanical properties on the grain size we generated polycrystalline samples with the average grain volumes of 16, 21, 30, 40, 50 and 64 $nm^3$, see Fig. \ref{fig:polycryst}.

In addition we have tested the convergence of mechanical properties with respect to the size of the simulation box for the same average grain size (namely $2\times2\times2$, $4\times4\times4$, and $8\times8\times8$ nm).

\subsection{\textit{Ab initio} calculations}

We used Density Functional Theory (DFT) as a first-principles method for training the MLIPs and validation of the results. DFT calculations were performed with the projector augmented-wave density functional theory (PAW-DFT) \cite{hohenberg1964inhomogeneous, kohn1965self} as implemented in the VASP package \cite{hafner2008ab, kresse1996efficient, kresse1993ab, kresse1994ab}.
The generalized gradient approximation with Perdew-Burke-Ernzerhof (GGA-PBE) \cite{perdew1996generalized} parametrization for exchange-correlation functional was used.
For each considered single crystal the PAW potentials were used according to the corresponding number of valence electrons to describe the electron-ion interactions.
The plane-wave energy cutoff of 500 eV and Methfessel-Paxton \cite{monkhorst1976special} smearing of electronic occupations ensured the convergence of total energies.
The $\Gamma$-centered $k$-point mesh of $8\times8\times8$ was used for Brillouin zone sampling .
For potential energy minimization we used a built-in conjugate gradient method with the maximum net force tolerance of less than 0.01 eV/\AA.

For initial training of a MLIP we actively selected atomistic configurations from \textit{ab initio} molecular dynamics. 
Timestep for AIMD was chosen to be equal to 1 fs.
The total time of each simulation was 2 ps.
The plane wave energy cutoff of 500 eV, the Methfessel–Paxton smearing \cite{monkhorst1976special} of electronic occupations, and $\Gamma$-centered $k$-point meshes with a resolution of $2\pi\times0.04 \AA^{-1}$ of the Brillouin zone sampling were used as implemented in VASP \cite{hafner2008ab, kresse1996efficient, kresse1993ab, kresse1994ab}. This ensures the convergence of the energy differences and stress tensors.
For more details about training procedure, calculation of MTP forces, readers are encouraged to check Ref. \cite{novikov2020mlip}. 

\subsection{Elastic moduli calculation}

The independent elastic constants for polycrystals were calculated following the standard atomistic simulation methodology as described in Ref. \cite{rassoulinejad2016evaluation}. %
This methodology involve 5 steps. 
\begin{enumerate}
    \item Structure relaxation.
    \item Applying a finite (about 1\%) positive and negative strain to the structure in all nonequivalent direction.
    \item Relaxation of the strained structure (with the fixed shape of the supercell).
    \item Calculation of the stresses for the strained structures.
    \item Calculation of the elastic constants using the stresses by finite differences.
\end{enumerate}
Elastic constants C relate the strain $\epsilon$ and the stress $\sigma$ in a linear fashion:
\begin{equation} \label{eq:elastic}
\sigma_{ij}  = \sum_{kl} C_{ijkl}\epsilon_{kl}
\end{equation}

For elastic tensor calculation the Large-scale Atomic/Molecular Massively Parallel Simulator (LAMMPS) package was used \cite{plimpton1995fast}. 

The values of elastic moduli have been calculated for different polycrystalline samples with the same average grain size and averaged. 
It should be noted that the generated samples are typically not isotropic and $C_{11}$$\neq$$C_{22}$$\neq$$C_{33}$, $C_{12}$$\neq$$C_{13}$$\neq$$C_{23}$, $C_{44}$$\neq$$C_{55}$$\neq$$C_{66}$. 
At the same time a polycrystalline diamond can be considered as isotropic at large scale. This fact allows us to consider $C_{22}$ and $C_{33}$ calculated for the same polycrystalline structure as additional sampled values for $C_{11}$. Similarly $C_{13}$ and $C_{23}$ are the sampled values for $C_{12}$, and $C_{55}$, $C_{66}$ are the sampled values of $C_{44}$. Thus elastic constants calculated for one polycrystalline structure yield 3 values of $C_{11}$, $C_{12}$ and $C_{44}$.

To control the statistical error we used k-means method with k=8. 
The statistical accumulation continued until sample variance of k-means were larger than 5\% of the average value.    

\subsection{MTP construction via active learning on-the-fly}

\begin{figure}
  \begin{center}
    \begin{minipage}[h]{1\linewidth}
      \center{\includegraphics[width=1\linewidth]{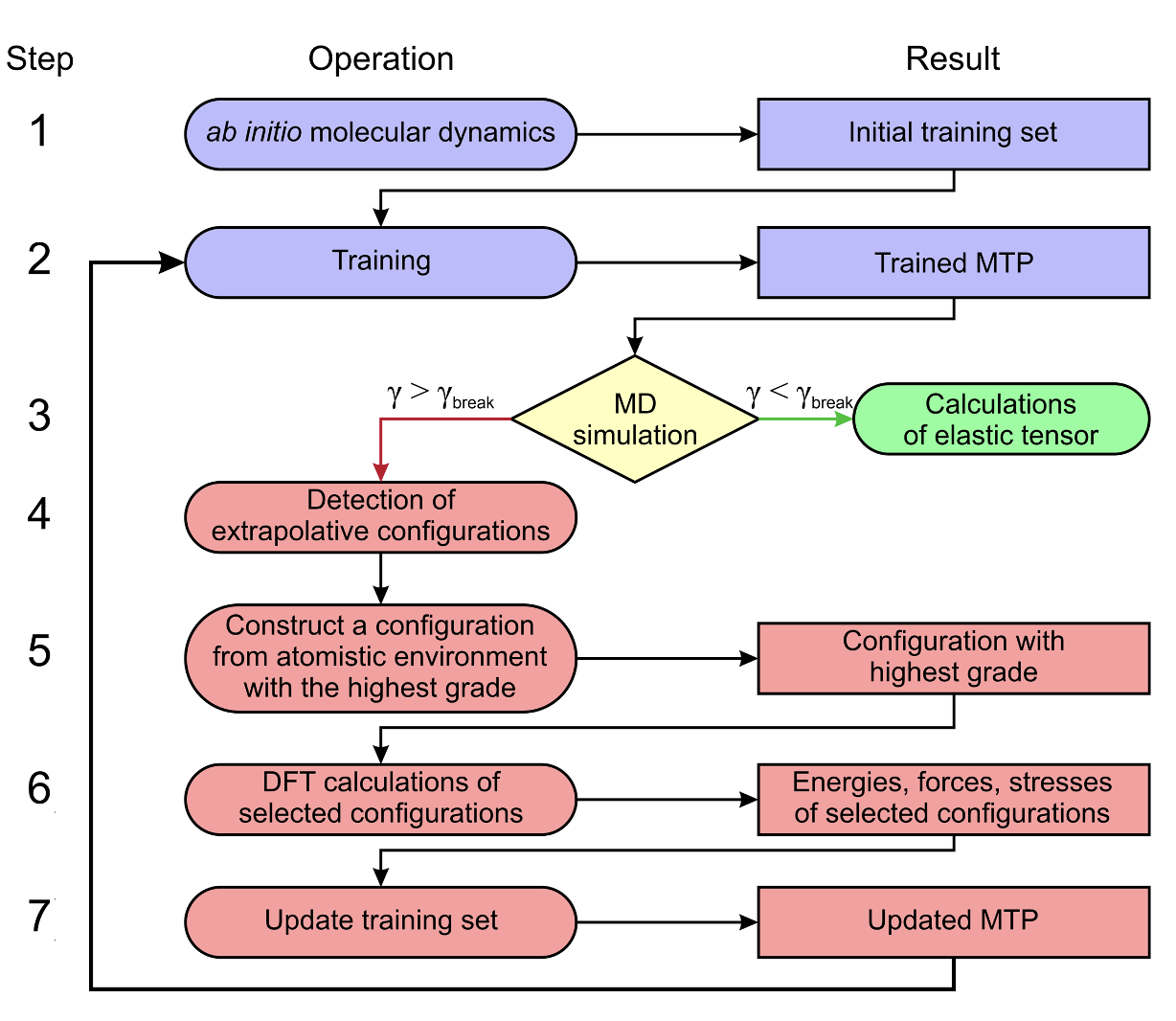}}  \\
    \end{minipage}
  \caption {Developed active learning bootstrapping iteration scheme for calculations of mechanical properties of crystalline and non-crystalline solids.}
  \label{fig:scheme}
  \end{center}
\end{figure}

Statistically reliable values of elastic moduli are averaged from the values calculated for dozens or even hundreds of samples. At the same time, calculation of elastic moduli for one sample assume a number of deformations and applying of relaxation procedure to the sample structure. Thus, the energy, forces and stresses are evaluated with MTP for atomistic configurations at each step of deformation-relaxation procedure. Some of these configurations may have local fragments where MTP extrapolate. In our scheme of active learning on-the-fly we evaluate the degree of extrapolation for each atomic environment in each configuration. If the extrpolation degree excceeds some critical value, the extrpolation fragment is processed with DFT and learned. This procedure is schematically shown in Figure\ \ref{fig:scheme}.

Active learning on-the-fly of MTP from scratch, however, is not computationally efficient. Therefore we pre-trained our MTP in passive manner with the atomistic configurations sampled from \textit{ab initio} molecular dynamics trajectories (step 1 in Fig.\ \ref{fig:scheme}). For this purpose we perform \textit{ab initio} molecular dynamics with DFT of 64 atoms of two-grain diamond over 1 ps (1000 timesteps).
  
After MTP training (step 2 in Fig.\ \ref{fig:scheme}) we started calculation of elastic tensor of studied system (Steps 3-7 in Fig.\ \ref{fig:scheme}).
This was performed with the active selection of extrapolative configurations, and the so-called one-extrapolation-threshold scheme (${\cal Y}_{break}$) was used \cite{novikov2020mlip}.
Exceeding ${\cal Y}_{break}$ indicates very high extrapolation degree and possible low accuracy in prediction of the energy, forces and stresses. 
Therefore we terminate elastic tensor calculation in order to retrain MLIP. 
These values ${\cal Y}_{break}$ was chosen to be 11 as providing according to our experience the optimal balance between the accuracy of MLIP and frequency of retraining. 
Detailed description of scheme is described in Ref.\ \cite{novikov2020mlip}.

After reaching the termination condition (${\cal Y \leq \cal Y}_{break}$) we select the configurations to be added to the training set among all extrapolative configurations for which the extrapolation was detected (step 4 in Fig.\ \ref{fig:scheme}).
Selection procedure is necessary to construct new active set from a pool of extrapolative configurations.

On the step 5 we extract from the large configuration a cubic box containing the local atomic environments causing the extrapolation.
The extrapolative atomic environment include the central atom and its neighborhood in the cutoff sphere which was taken of 5 \AA. 

Constructed local atomic structure extracted from polycrystalline typically contains around 100 atoms.
At the next step (step 6) this atomic configuration expands to a periodic structure with the relaxation on DFT of atoms outside the extrapolative environment in order to minimize energy of the periodic interface. 
At the same step the DFT calculations of energy, forces, and stresses is performed.
Further steps of adding to the initial training set with subsequent retraining of MTP (steps 7 and 2 in Fig. \ref{fig:scheme}).
By using this scheme there are no needs to consider entire polycrystalline structure in the DFT calculations in order to actively learn MTP.
Thus, as the first iteration of active learning of MTP was finished and MD simulation of elastic tensor can be continued with updated actively learned MTP until the critical value of extrapolation is reached again or calculation of all configuration is finished.
Each iteration of this scheme expands the training domain and improves the transferability of the MTP (i.e., the amount of extrapolations and the extrapolation degree is reduced).
As was discussed above, similar approach was recently used in Ref. \cite{podryabinkin2022nanohardness} for simulation of nanohardness of single crystal compounds and in MD run for copper\cite{podryabinkin2023mlip}.

\begin{figure*}
\begin{center}
   \centering
	\includegraphics[width=1\textwidth]{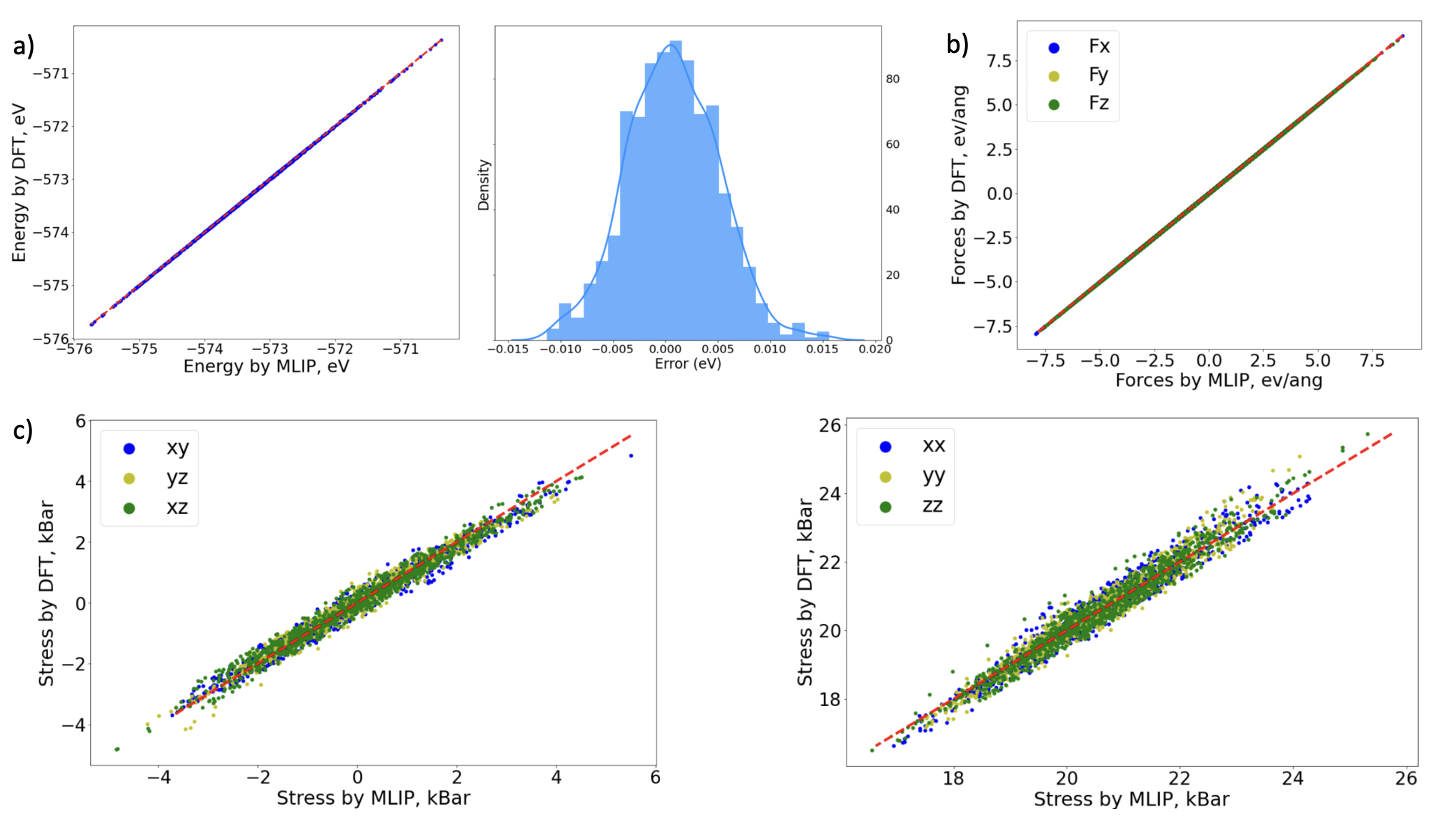}
\caption{Calculated by DFT and fitted by MLIP values of (a) total energy with error distribution, (b) forces, and (c) stresses, obtained for MTP for single crystal diamond.}
\label{fig:diamond_mono_dft_mlip}
\end{center}
\end{figure*}

The MTPs for single crystals was applied for diamond, Si, SiC, WC, and CrB$_4$. 
Detailed information about obtained results for studied single crystals is shown in Tables S1-S6 in {Supporting Information}. 

%%%%%%%%%%%%%%%%%%%%%%%%%%
 
%%%%%%%%%%%%%%%%%%%%%%%%%%
\section{3. Results and Discussion}
\label{sec:result}
\subsection{MTP for polycrystalline diamond}

The accuracy of obtained MTP for single crystal diamond, as the base for learning of MTP for polycrystals was estimated.
Calculated by DFT total energies, forces, stresses and their fitted quantities by MTP for single crystal diamond are presented in Fig. \ref{fig:diamond_mono_dft_mlip}.
All metrics are presented for every configuration in training set.
For calculated and fitted energies (Fig. \ref{fig:diamond_mono_dft_mlip}a) the maximal absolute difference is $6.7\times10^{-2}$ eV, average absolute difference is $4.1\times10^{-3}$ eV, and RMS absolute difference is $7.1\times10^{-3}$ eV.
Error distribution shows the relation between calculated and fitted energies.
It is highly symmetrical around zero and might be considered as Gaussian type.
From this fact we can conclude that MTPs have no systematic bias towards the overestimation and underestimation of results.

For calculated and fitted forces (see Fig. \ref{fig:diamond_mono_dft_mlip}b) maximal absolute difference, average absolute difference, and RMS absolute difference are 1.3 eV/\AA,
$2\times10^{-2}$ eV/\AA, and $2.2\times10^{-2}$ eV/\AA respectively.
For stresses we obtained the following values of 2.5, 0.7 and 0.7 kBar for maximal absolute difference, average absolute difference, and RMS absolute difference, respectively, see Fig. \ref{fig:diamond_mono_dft_mlip}c.
All obtained trend lines and calculated absolute differences for energies, forces and stresses can interpret an accurate predictive power of used MLIP.

\begin{figure*}
\begin{center}
   \centering
	\includegraphics[width=1\textwidth]{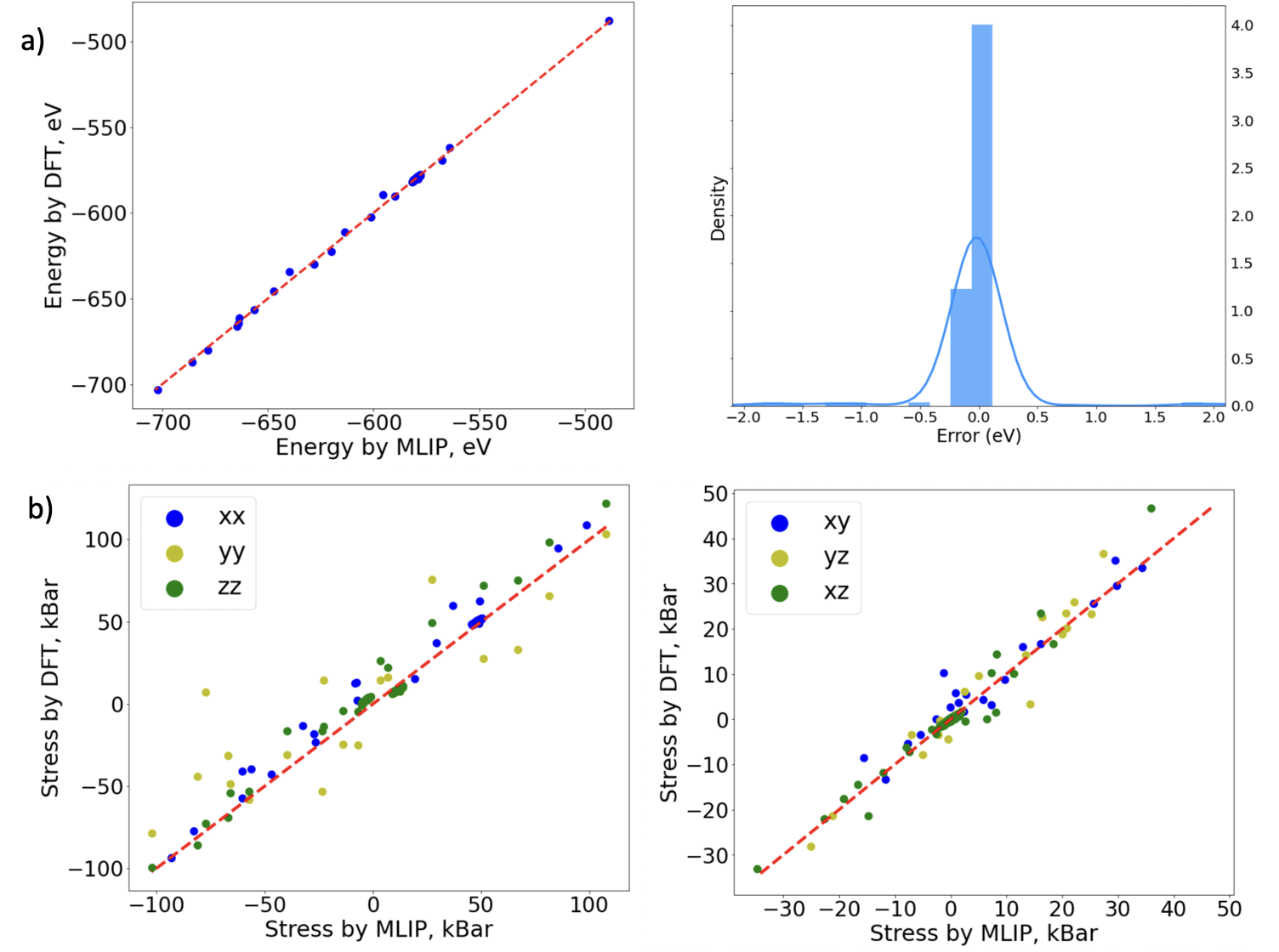}
\caption{Calculated by DFT and fitted by MLIP values of (a) total energy with
error distribution and (b) stresses only for local configurations extracted from the polycrystal for active learning.}
\label{fig:diamond_poly_en_3}
\end{center}
\end{figure*}

The accuracy of actively learnt MTP on local atomistic environment for polycrystalline diamond was also estimated.
Calculated by DFT total energies, forces, stresses and fitted by MLIP only for local configurations extracted from polycrystal are presented in Fig. \ref{fig:diamond_poly_en_3}.
For calculated and fitted energies (Fig. \ref{fig:diamond_poly_en_3}a) the maximal absolute difference is $7.6\times10^{-2}$ eV, average absolute difference is $1.9\times10^{-3}$ eV, and RMS absolute difference is $7.9\times10^{-3}$ eV.
Error distribution shows the relation between DFT and MTP energies.

For stresses we obtained the values of 10.7, 2.7 and 3.1 kBar for maximal absolute difference, average absolute difference, and RMS absolute difference, respectively, see Fig. \ref{fig:diamond_poly_en_3}b.
All obtained trend lines and calculated absolute differences for energies, forces and stresses demonstrate an accurate predictive power of used MLIP.

\begin{figure}
\begin{center}
   \centering	\includegraphics[width=0.45\textwidth]{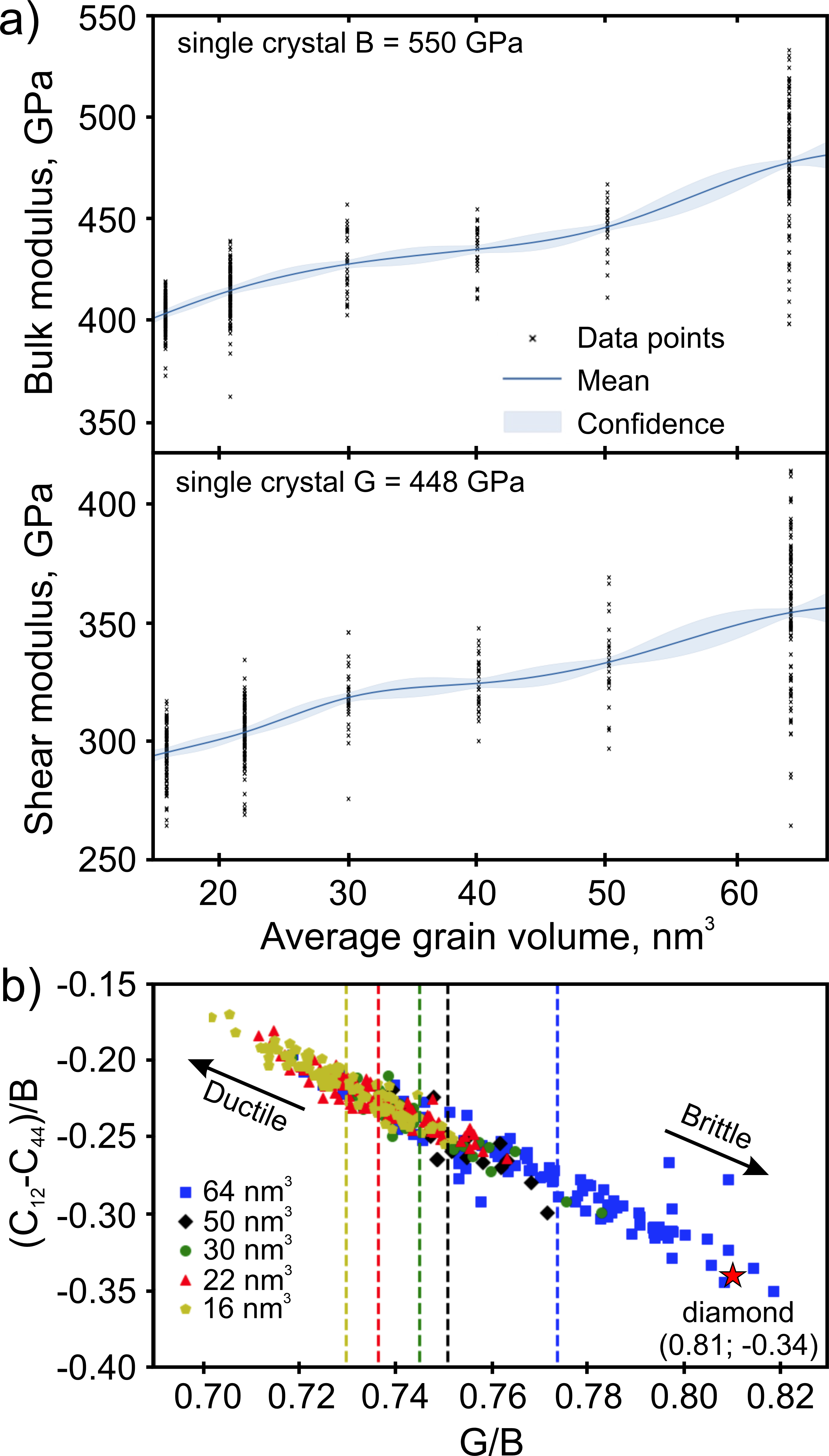}
\caption{a) Dependence of the bulk (top) and shear (bottom) moduli of diamond polycrystals with different grain sizes in comparison with single crystal diamond. b) Correlation between (C$_{12}$-C$_{44}$)/B and $G/B$, where B is bulk modulus, G is shear modulus showing the changing in ductility and brittleness of polycrystals depending on the grain size. Dashed vertical lines denote the average values among structures with same grain size. Value for diamond is shown by red star.}
\label{fig:diamond_poly_gaus}
\end{center}
\end{figure}

\subsection{Mechanical properties of polycrystalline diamond}

Polycrystals can be considered as orthotropic materials where 9 independent second order elastic constants are presented and should be calculated, namely $C_{11}$, $C_{22}$, $C_{33}$, $C_{44}$, $C_{55}$, $C_{66}$, $C_{12}$, $C_{13}$, and $C_{23}$. 
By using the combination of  these components of elastic tensor the elastic moduli were determined via the Voigt-Reuss-Hill averaging.
The results of calculations of elastic moduli of polycrystalline diamond with different grain size by using actively learned MTP on local environments are shown in Fig.~\ref{fig:diamond_poly_gaus}.
One can see that bulk modulus of polycrystalline diamond increases with increasing the average size of the grains tending to the bulk modulus of single crystal diamond as limiting case (horizontal dashed purple line in Fig.~\ref{fig:diamond_poly_gaus}).
For each grain size number of structures (from 23 to 100) were generated explaining the deviation of calculated bulk modulus.
For each grain size the sample variance $S$ and sample mean $M$ values were calculated.
We continue generating and calculating elastic moduli until the statistical error become less than 1\%.

The average size of grains for which the diamond polycrystals were generated was selected by using Gaussian process (GP) with the radial basis function (RBF) kernel.
Initially we simulated 2 diamond polycrystals with average grains size 16 and 64 $nm^3$ and calculated elastic constants for them according to our setup (Fig. \ref{fig:scheme}).
Results of bulk modulus for these two polycrystals are shown in Fig. S2 in Supporting Information. 
According to these results we determined the confidence parameters in GP to define the grain size for further polycrystals in order to minimize the confidence parameters, see Fig. S2 in Supporting Information.
Then the other sizes, namely 40, 30, 50, and 22 $nm^3$ (in this order) were added for consideration to minimize the confidence parameter in GP (grey area in the Figure~\ref{fig:diamond_poly_gaus}).

Obtained results on the bulk moduli of considered polycrystals show monotonic growth starting from 400 GPa (near to amorphous carbon structure) to 480 GPa for structure with average grain volume of 64 nm$^3$, see Figure~\ref{fig:diamond_poly_gaus}.
The average value of the bulk modulus for polycrystal with largest grain size is about 480 GPa, which is under to calculated value for single crystal diamond of 550 GPa and is within the confidence interval of our calculations. 

To understand how the grain size influence the ductility and brittle behavior of polycrystals we have calculated the Pugh-Pettifor\cite{senkov2021generalization} criterion as shown in Fig. \ref{fig:diamond_poly_gaus}b. 
Correlation between (C$_{12}$-C$_{44}$)/B and G/B allows us to determine the ductility and brittleness of polycrystals.
As one can see the polycrystals with small grain size (16, 22 $nm^3$) are more ductile compared to larger grain sizes, see Fig. \ref{fig:diamond_poly_gaus}b.
As the grain size increases the polycrystals become more brittle.
The average G/B ratio for polycrystals with grain of 64 $nm^3$ is 0.775 and the maximum value is about 0.82, see Fig. \ref{fig:diamond_poly_gaus}b.
According to this data mechanical stiffness of considered polycrystals does not exceed the value of single crystal diamond (G/B is 0.81).
Thus, all considered polycrystals with various grain sizes are less brittle according to Pugh-Pettifor\cite{senkov2021generalization} criterion compared to single crystal diamond. 

%%%%%%%%%%%%%%%%%%%%%%%%%%
\section{Conclusion}
\label{sec:conclusion}
We have developed the active learning bootstrapping iteration scheme for calculations of elastic tensor of complex solids, namely composites, polycrystals, and multiphase systems, by using machine learning interatomic potentials with active learning on local atomic environments.
Our scheme allows one to achieve high accuracy in simulating the elastic properties of complex solids.
The proposed scheme was used to calculate the elastic tensor and elastic moduli both for single crystals with various structures and compositions and polycrystalline structure.
To evaluate our approach, diamond polycrystals were assessed, and the resulting elastic properties were compared to existing reference data, demonstrating excellent conformity and precision.
Developed approach allows one to study mechanical properties of materials that usually synthesized and used in experiments, i.e. noncrystalline materials.
This enables comprehensive investigations into the mechanical properties of complex materials, such as polycrystals and composites, bringing the obtained data closer to that found in experiments.

\section{Competing Interests}
The Authors declare no Competing Financial or Non-Financial Interests

\begin{acknowledgements}
This work was carried out using our Oleg supercomputer of Computational Materials Discovery Laboratory and the \textit{ElGatito} and \textit{LaGatita} supercomputers of the Industry-Oriented Computational Discovery group at the Skoltech Project Center for energy Transition and ESG.
\end{acknowledgements}

\bibliography{main}

\end{document}

\title{Supporting Information for Mechanical properties of single and polycrystalline solids from machine learning}

\author{Faridun N. Jalolov}
\affiliation{Skolkovo Institute of Science and Technology, Skolkovo Innovation Center, Bolshoy Boulevard 30, bld. 1, Moscow 121205, Russia}

\author{Evgeny V. Podryabinkin}
\affiliation{Skolkovo Institute of Science and Technology, Skolkovo Innovation Center, Bolshoy Boulevard 30, bld. 1, Moscow 121205, Russia}

\author{Artem R. Oganov}
\affiliation{Skolkovo Institute of Science and Technology, Skolkovo Innovation Center, Bolshoy Boulevard 30, bld. 1, Moscow 121205, Russia}

\author{Alexander V. Shapeev}
\affiliation{Skolkovo Institute of Science and Technology, Skolkovo Innovation Center, Bolshoy Boulevard 30, bld. 1, Moscow 121205, Russia}

\author{Alexander G. Kvashnin*}
\affiliation{Skolkovo Institute of Science and Technology, Skolkovo Innovation Center, Bolshoy Boulevard 30, bld. 1, Moscow 121205, Russia}

\email{A.Kvashnin@skoltech.ru}

\date{\today}

\maketitle

\section{Geometry relaxation of single crystals using DFT}

The geometry of the primitive cells of the considered single crystals was taken from Materials Project database \cite{jain2013commentary}.
These structures were then relaxed and total energy computations were performed with the projector augmented-wave density functional theory (PAW-DFT) \cite{hohenberg1964inhomogeneous, kohn1965self} as implemented in the VASP package \cite{hafner2008ab, kresse1996efficient, kresse1993ab, kresse1994ab}.
The generalized gradient approximation of Perdew - Burke - Ernzerhof (GGA-PBE) \cite{perdew1996generalized} was used for the exchange - correlation functional.
For the diamond single crystal the PAW potentials were used according to its valence electrons to describe the electron-ion interactions.
Crystal structure was relaxed until the maximum net force on atoms became less than 0.01 eV/\AA.
The plane wave energy cutoff of 500 eV and Methfessel-Paxton \cite{methfessel1989high} smearing of electronic occupations ensured the convergence of total energies.
The $\Gamma$- centered $k$-point meshes of $8\times8\times8$ were used for Brillouin zone sampling.

\section{Additional data on diamond polycrystals}

\begin{figure}[h!]
  \begin{center}
    \includegraphics[width=1\textwidth]{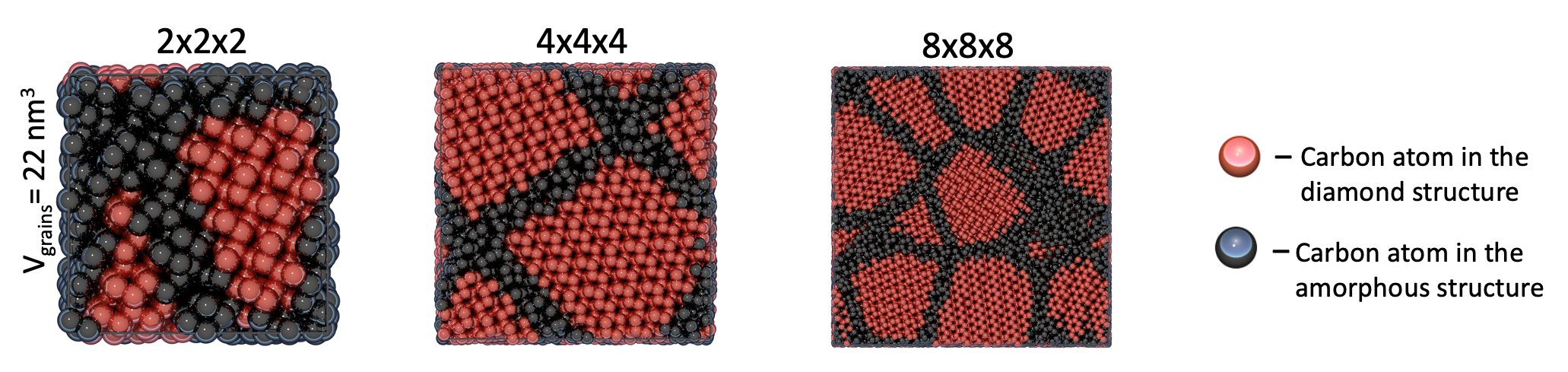}
    \caption{Atomic structure of slices of generated polycrystals with a different size of simulation boxes containing of 4 grains with average volume of each of $22$ $nm^3$. Red regions present regions with crystalline diamond structure, while black ones represent amorphous grain boundaries. The crystal structures were visualized using Ovito Pro software \cite{stukowski2009visualization}.}
\label{fig:poly}
  \end{center}
\end{figure}

\section{Elastic moduli of single crystal diamond}

The results of calculated elastic tensor for single crystal diamond obtained by using DFT, classical Tersoff potential, and MLIP without and with active learning are presented in Table S\ref{table_1}.   
It turns out that MLIPs with active learning can optimize data and hence a potential.
One can see good agreement between MLIP data and experimental values from Ref. \cite{mcskimin1972elastic}.
During the active learning of this MTP for diamond, it was retrained 3 times requiring only 3 additional single-point DFT calculations in order to determine interatomic forces, stresses, and energy of the selected configuration on which the extrapolation occurs.
Elastic constant $C_{11}$ determined by MTP is 1071 GPa and agrees well with reference experimental data of 1079 GPa \cite{mcskimin1972elastic}, while Tersoff potential underestimates this value (1056 GPa),  Table S\ref{table_1}.
Considering elastic constant $C_{12}$ one can observe little overestimation by MLIP in comparison with both DFT and experimental values (144 GPa in comparison with 129 GPa and 124 GPa \cite{mcskimin1972elastic} respectively).
Elastic constant $C_{44}$ calculated by MTP agrees much better with experiment in comparison with Tersoff potential.

\begin{table}
    \centering
     \caption{Calculated elastic constants and elastic moduli of single diamond crystal by using MLIP with passive learning and active learning (AL) in comparison with DFT, Tersoff empirical potential and referenced experimental data}
    \begin{tabular}{ |p{2.5cm}||p{2cm}|p{2cm}|p{2cm}| p{2cm}|p{2cm}| | p{2cm}| }
 \hline
 Method & DFT (PAW-PBE) & Tersoff pot.& MLIP (passive learning) & MLIP with AL & Exp\cite{mcskimin1972elastic}\\
 \hline
 $C_{11}$, GPa & 1054 & 1056 & 1086 & 1071 & 1079\\
 $C_{12}$, GPa   & 129 & 107 &  271 & 144 & 124\\
 $C_{44}$, GPa  & 559  & 635 &  646 & 558 & 578\\
 B, GPa & 437  &  423 & 542 & 453 & 442  \\
 G, GPa & 520 &  571  & 550 & 520 &  537 \\
 Y, GPa & 1118 &  1181 & 1234& 1129 & 1148 \\
  \hline
 \end{tabular}

    \label{table_1}
\end{table}

In order to show the performance in accuracy of active learning scheme we have also calculated elastic tensor of single crystal diamond by using passively learned MTP as shown in the table. 
We have also calculated the elastic moduli, namely bulk modulus, shear modulus, and Young's modulus.
As one can see calculated elastic moduli by MTP perfectly agree with reference experimental data.
Young's modulus of single crystal diamond was calculated to be 1129 GPa which is almost the same for other methods. 
For bulk and shear moduli the results obtained by using MTP closer to experimental data rather than values obtained by other methods.

\section{Elastic properties of covalent compounds}

In table \ref{table_si} the calculated elastic properties for bulk silicon in comparison with experimental result \cite{norton1973impurity} are shown. 
The single crystal system contained 64 atoms. 
For silicon single crystals the elastic properties were calculated by using different methods, namely the DFT, Tersoff interatomic potential and MLIPs with active learning. 
Due to silicon crystals structure has high symmetry, there are only three elastic constants($C_{11}$, $C_{12}$, and $C_{44}$). 
By using these three independent elastic constants we calculated the elastic moduli by using the Voight-Reuss-Hill average.

The calculated elastic properties of silicon calculated by DFT and MLIPs are in a good correlation  with each other. 
The differences between elastic constants do not exceeds $4$ GPa (around 2\%). 
The experimental results are a bit overestimated by the calculated results. 
Nevertheless, the differences of elastic constants and moduli are not more than 6\%  in comparison with experiment. 

The calculated elastic properties of another covalent binary material $c-SiC$ are shown in table \ref{table_csic}.
For this material the elastic properties were also calculated by using different methods (DFT, CMD by using Tersoff potentials, and MLIPs with active learning).
The calculated elastic constants and elastic moduli by DFT and MLIPs are correlated well with each other, which is expected.
The differences between these two methods do not exceed 1\%. 
Moreover, the result obtained with MLIP and DFT is in a good correlation with referenced experimental results \cite{levinshtein2001properties}.
The experimental results do not overestimate the calculated data and the differences of calculated elastic constants and moduli compared to experimental data are around 2\%.

Overall, the MTPs are able to calculate the elastic properties of the cubic system with first-principles accuracy.
MTPs were trained on $ab$ $initio$ MD trajectories, and the accuracy of elastic properties calculation with DFT and MTPs was expected to be similar. 
Moreover, MTPs are shown better results rather than Tersoff empirical potentials for materials with FCC lattice.

Another considered material is $4h-SiC$ (hexagonal polymorphic modification of SiC).
The calculated elastic constants and moduli by using different methods in comparison with experimental results are shown in table \ref{table_hsic}.
Due to hexagonal symmetry of $4h-SiC$ there are five independent elastic constants ($C_{11}$, $C_{12}$, $C_{13}$, $C_{33}$ and $C_{44}$) should be calculated.
The combination of these elastic constants was used to determine the elastic moduli by using the Voigt-Reuss-Hill averaging.

Comparison of calculated elastic properties by MTPs for $4h-SiC$ with experimental data from Ref. \cite{nawaz2017nano} shows good agreement.
The  differences between DFT results and MLIPs do not exceed 10 GPa, which is not more  than 3\%.
The experimental results are a bit higher than the calculated ones with DFT and MLIPs.
Nevertheless, the differences in elastic constants and moduli are not more than 5\% and 2\% respectively.

The calculated elastic properties of  single crystal of $WC$ are shown in table \ref{table_wc}).
The elastic constants were calculated by using DFT and MLIPs (there are no empirical potentials available for description of W-C interactions), and the results of the calculations agree well with experimental results from Ref. \cite{sung1997reactivities}.
This indicates that the MLIPs with active learning are able to describe also hexagonal binary single crystals with comparable accuracy to DFT.

For the $CrB_4$ single crystal ($Pnnm$ space group), which is one of the most hardest transition metal borides, we have also trained MLIPs with active learning technique.
This compound has the most complicated crystal structure compared to previously described.
It should be noted, that there are no empirical potentials available for such compound.
Due to low symmetry of $CrB_4$, there are 9 independent elastic constants ($C_{11}$, $C_{22}$, $C_{33}$, $C_{44}$, $C_{55}$, $C_{66}$, $C_{12}$, $C_{13}$, and $C_{23}$).
Calculated 9 independent elastic constants were used for determination of the elastic moduli by using Voight-Reuss-Hill averaging.
The results are shown in table \ref{table_crb4}.

The differences of calculated elastic constants by MLIPs and DFT in comparison with experimental results \cite{wang2014crystal} are around 20\% (the highest differences observed for $C_{66}$). 
Nevertheless, the calculation of elastic moduli shows that the differences of elastic moduli do not exceed 5\%, which indicates the ability of actively learned MLIPs to describe unary and binary compounds with crystal structure of various complexity.

\begin{table}
    \centering
     \caption{Calculated elastic constants and elastic moduli of the cubic Si by using MLIP with active learning (AL) in comparison with DFT, Tersoff empirical potential and referenced experimental data}
    \begin{tabular}{ |p{2.5cm}||p{2cm}|p{2cm}|p{2cm}|p{2cm}| | p{2cm}| }
 \hline
 Method & DFT (PAW-PBE) & Tersoff pot. & MLIP with AL & Exp\cite{mcskimin1972elastic}\\
 \hline
 $C_{11}$, GPa & 153 & 140 & 158 & 166 \\
 $C_{12}$, GPa   & 73 & 75 &  77 & 80 \\
 $C_{44}$, GPa  & 57  & 66 &  56 & 64 \\
 B, GPa & 100  &  97 & 104 & 108  \\
 G, GPa & 50 &  53  & 50 & 56  \\
 Y, GPa & 129 &  132 & 128 & 142  \\
 \hline
 \end{tabular}
    \label{table_si}
\end{table}

\begin{table}
    \centering
     \caption{Calculated elastic constants and elastic moduli of the cubic SiC by using MLIP with active learning (AL) in comparison with DFT, Tersoff empirical potential and referenced experimental data}
    \begin{tabular}{ |p{2.5cm}||p{2cm}|p{2cm}|p{2cm}|p{2cm}| | p{2cm}| }
 \hline
 Method & DFT (PAW-PBE) & Tersoff pot. & MLIP with AL & Exp\cite{levinshtein2001properties}\\
 \hline
 $C_{11}$, GPa & 383 & 413 & 380 & 382 \\
 $C_{12}$, GPa   & 128 & 113 &  125 & 126 \\
 $C_{44}$, GPa  & 235  & 232 &  241 & 241 \\
 B, GPa & 213  &  215 & 210 & 211  \\
 G, GPa & 192 &  214  & 196 & 195  \\
 Y, GPa & 443 &  481 & 448 & 447  \\
 \hline
 \end{tabular}

    \label{table_csic}
\end{table}

\begin{table}
    \centering
     \caption{Calculated elastic constants and elastic moduli of the hexagonal SiC by using MLIP with active learning (AL) in comparison with DFT, Tersoff empirical potential and referenced experimental data}
    \begin{tabular}{ |p{2.5cm}||p{2cm}|p{2cm}|p{2cm}|p{2cm}| }
 \hline
 Method & DFT (PAW-PBE) & MLIP with AL & Exp\cite{nawaz2017nano}\\
 \hline
 $C_{11}$, GPa & 486 & 497 & 501\\
 $C_{12}$, GPa   & 104 & 96 &  111 \\
 $C_{13}$, GPa   & 51 & 53 &  52  \\
 $C_{33}$, GPa  & 535  & 530 &  553  \\
 $C_{44}$, GPa  & 160  & 160 &  163  \\
 B, GPa & 108  &  107 & 110  \\
 G, GPa & 91 &  92  & 94   \\
 Y, GPa & 617 &  624 & 634   \\
 \hline
 \end{tabular}

    \label{table_hsic}
\end{table}

\begin{table}
    \centering
     \caption{Calculated elastic constants and elastic moduli of the hexagonal WC by using MLIP with active learning (AL) in comparison with DFT, Tersoff empirical potential and referenced experimental data}
    \begin{tabular}{ |p{2.5cm}||p{2cm}|p{2cm}|p{2cm}|p{2cm}| }
 \hline
 Method & DFT (PAW-PBE) & MLIP with AL & Exp\cite{sung1997reactivities}\\
 \hline
 $C_{11}$, GPa & 729 & 681 & 699\\
 $C_{12}$, GPa   & 212 & 211 &  226 \\
 $C_{13}$, GPa   & 169 & 178 &  174  \\
 $C_{33}$, GPa  & 889  & 826 &  938  \\
 $C_{44}$, GPa  & 258  & 235 &  237  \\
 B, GPa & 196  &  188 & 199  \\
 G, GPa & 133 &  123  & 121   \\
 Y, GPa & 881 &  802 & 833   \\
 \hline
 \end{tabular}

    \label{table_wc}
\end{table}

\begin{table}
    \centering
     \caption{Calculated elastic constants and elastic moduli of the orthorhombic $CrB_4$  by using MLIP with active learning (AL) in comparison with DFT, Tersoff empirical potential and referenced experimental data}
    \begin{tabular}{ |p{2.5cm}||p{2cm}|p{2cm}|p{2cm}|p{2cm}| p{2cm}| }
 \hline
 Method & DFT (PAW-PBE) & MLIP with AL & Exp\cite{wang2014crystal} & DFT \cite{} \\
 \hline
 $C_{11}$, GPa & 555 & 556 & 554 & 542\\
 $C_{22}$, GPa & 880 & 848 & 880 & 855\\
 $C_{33}$, GPa  & 502  & 524 &  473  & 492\\
 $C_{44}$, GPa  & 255  & 215 &  254 & 252\\
 $C_{55}$, GPa  & 282  & 243 &  282 & 280\\
 $C_{66}$, GPa  & 241  & 195 &  250 & 253\\
 $C_{12}$, GPa   & 72 & 63 &  65 & 50\\
 $C_{13}$, GPa   & 100 & 92 &  107 & 104 \\
 $C_{23}$, GPa   & 90 & 72 &  95 & 87 \\

 B, GPa & 265  &  264 & 265 &  251\\
 G, GPa & 267 &  244  & 263  &  269\\
 Y, GPa & 834 &  810 & 825 &  840\\
 \hline
 \end{tabular}

    \label{table_crb4}
\end{table}

\section{Elastic constant at different temperatures by Tersoff potential}

It should be noted that classical molecular dynamics (CMD) simulation attempts to measure small differences in statistical averages generated by finite temperature.
This creates many opportunities for errors to occur, compared to zero-temperature calculation.
Hence, we calculated elastic constants for single crystal diamond using classical molecular mechanics (at zero temperature) and classical molecular dynamics (at finite temperature) simulations.
The results of calculations at finite temperature (300 K) and zero temperature are in good agreement with each other (see Table \ref{tab:diam-elast}).
The elastic constants at 0 K temperature is higher a bit, which are expected. 

\begin{table}[h!]
  \centering
      \caption{Comparative analysis of elastic constants for single crystal diamond at finite and zero temperatures by Tersoff potential}
    \begin{tabular}{||c | c | c | c ||} 
    \hline
    \hline
 Elastic constants & $T = 0$ K & $T = 300$ K & $T =300$ K Ref.\ \cite{mcskimin1972elastic} \\
    \hline
    \hline
 $C_{11}$ (GPa)   & 1074  & 1056 & 1079 \\
 $C_{12}$ (GPa)   & 102   & 107 & 124 \\
 $C_{44}$ (GPa)   & 641   & 635 & 578 \\
    \hline
    \end{tabular}
    \label{tab:diam-elast}
\end{table}

\begin{figure}[h!]
  \begin{center}
    \includegraphics[width=1\textwidth]{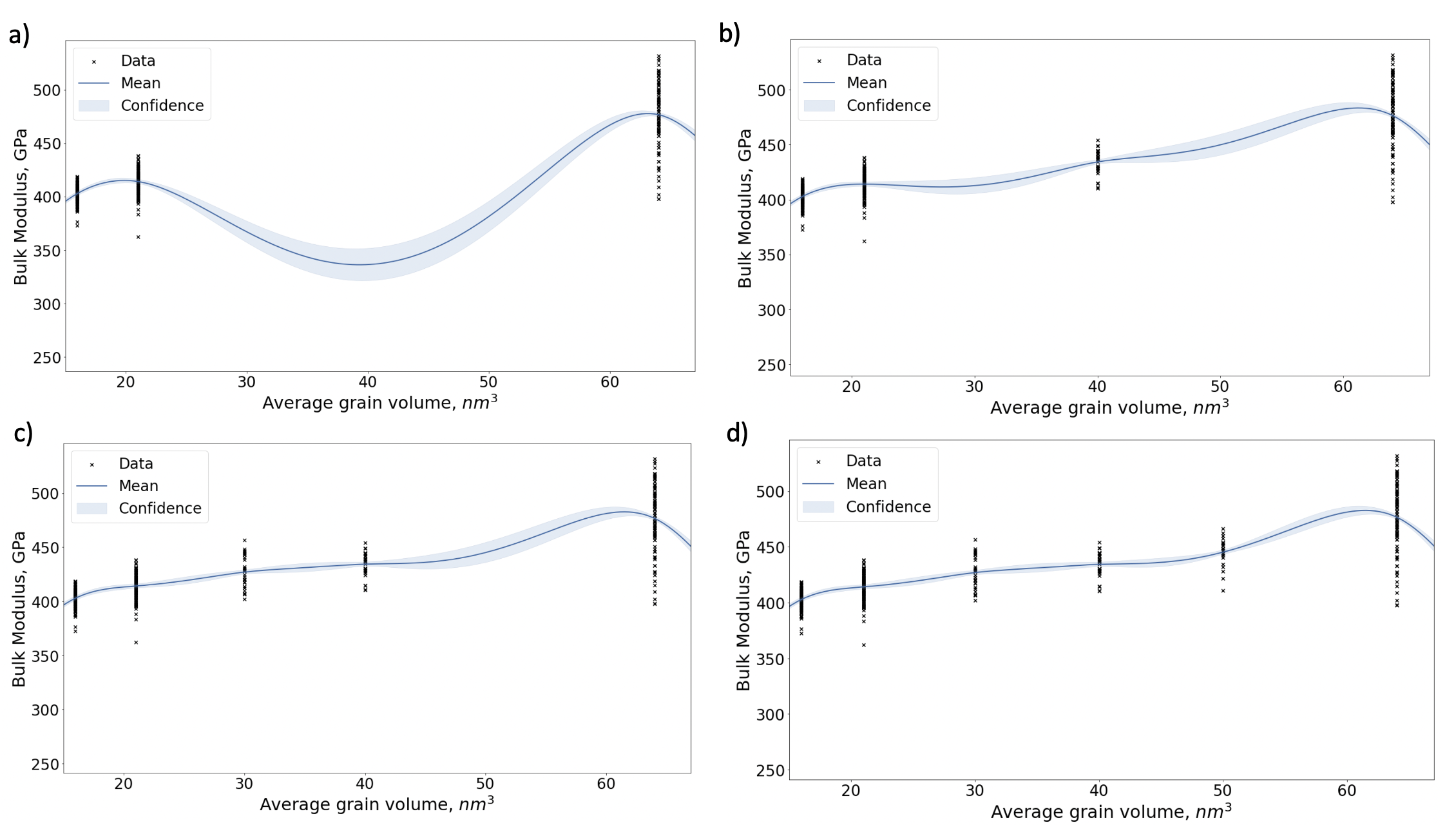}
    \caption{Optimization bulk modulus dependence on grain size of polycrystals with Gaussian process}
\label{fig:b}
  \end{center}
\end{figure}

\clearpage
%%%%%%%%%%%%%%%%%%%%%%%%%%
%\medskip
%\printbibliography
\bibliography{main}